\documentclass{aastex}

\newcommand{\emaila}{ E-mail: mfullana@mat.upv.es}
\newcommand{\emailb}{ E-mail: aalfonsofaus@yahoo.es}

\begin{document}
\title{Two restrictions in the theories that include $G(t)$ and $c(t)$ varying with time}


\author{M\`arius Josep Fullana i Alfonso}
\affil{Institut de Matem\`atica Multidisciplin\`aria, 
Universitat Polit\`ecnica de Val\`encia, 
Cam\'{\i} de Vera, 
Val\`encia, 46022, Spain.\\
\emaila}
\and
\author{Antonio Alfonso-Faus}
\affil{Escuela T\'ecnica Superior de Ingenier\'{\i}a Aeron\'autica y del Espacio, 
Plaza del Cardenal Cisneros, 3,
Madrid, 28040, Spain.\\
\emailb}

\begin{abstract}
Much work has been done taking into account the possibility that the gravitational {\it constant} $G$
may vary with cosmological time $t$ (or with the cosmological scale factor $a(t)$). The same may be said about the speed of light $c$. We present here two
important remarks on these subject. These remarks include $G(t)$ and $c(t)$ varying with time
with the restriction $8 \pi G / c^4 = \hbox{constant}$.
\end{abstract}

\keywords{Cosmology; gravity; constants.}

\section{Introduction}	

Much work has been done taking into account the possibility that the gravitational {\it constant} 
$G$ may vary with cosmological time $t$ (or with the cosmological scale factor $a(t)$). In many of these works the speed of light $c$ is not addressed or is taken as constant (see for example the Dirac large number hypothesis, 
\cite{Dirac}). 
The same can be said of many works that assume the speed of light $c$ to vary with cosmological time $t$ or $a(t)$ (see for example 
\cite{ MaMo}). Again in many of these works the gravitational constant $G$ is not addressed or is taken as constant. 
We have two important remarks on these assumptions.

\section{First remark}

As shown below, both $G$ and $c$ must be related by the condition:

\begin{equation}
8 \pi G / c^4 = \hbox{constant}
\label{e1}
\end{equation}

It is well known that the covariant divergence of the first member of the Einstein field equations (the Einstein tensor $G^{\mu \nu}$ including the 
$\Lambda$ constant term) is identically zero, a mathematical result. The field equations are

\begin{equation}
R_{\mu \nu} - \frac{1}{2} g_{\mu \nu} R + g_{\mu \nu} \Lambda = \frac{ 8 \pi G}{c^4} T_{\mu \nu}
\label{eF}
\end{equation}

Then, the covariant divergence of the second member of (\ref{eF}) must also be zero. The resultant equation is therefore

\begin{equation}
\nabla \left ( 8 \pi G/c^4 \right ) T^{\mu \nu}  = 0
\label{e2}
\end{equation}

\noindent
where $T^{\mu \nu}$ is the stress-energy tensor. For $G$ (the gravitational constant) and $c$ (the speed of light) taken as constants 
we get the usual local conservation of energy and momentum from the relations

\begin{equation}
(8 \pi G/c^4 = \hbox{constant} ) , \ \ \ \ \nabla T^{\mu \nu}  = 0
\label{e3}
\end{equation}

One gets the energy and momentum conservation laws that are experimentally so well validated to date. 
We remark now that it is not necessary that both $G$ and $c$ be constants to get (\ref{e3}). 
It is enough to impose the constancy given in equation (\ref{e1}) to get the same relation (\ref{e3}), and thus maintaining the validity of the conservation laws of physics. 
Then, it is clear that if we allow $G$ to vary with time one has to take into account that $c$ must also vary with time, 
according to equation (\ref{e1}), and vice versa. In a system of units so defined that $G = c^4$ special relativity factors like $c^2$ may be replaced by $G^{1/2}$.

This new possibility allows us to point out the following: 
it may represent an open door towards the finding of a linkage between special relativity and gravitation. 
In particular, the gravitational radius $r_g$ of a mass $M$, the Schwarzschild radius that is of the order $GM/c^2$, 
is here (with $G = c^4$) equal to the relativistic energy $Mc^2$. And since the gravitational radius is a length, 
we have here the parallel physical condition between a size (a one dimensional brane, a string that incorporates gravity and quantum states) 
and energy, a quantized physical entity

\begin{equation}
r_g \approx M c^2 \ \ \ \hbox{and} \ \ \ (r_g)^2 \approx (Mc^2)^2 = GM^2
\label{eQ}
\end{equation}

The two dimensional brane, $(r_g)^2$ is seen here to be proportional to the self gravitational potential energy of the mass $M$, $GM^2$.

\section{Second remark}

{\it Theories that include time variations of the constants G and/or c must  take into account that both constants must be allowed to vary with time, restricted by the condition $8 \pi G/c^4 = constant$. If one needs to use the Einstein cosmological equations, they have to be derived including the two conditions, 
$G(t)$ and $c(t)$ with such a restriction, from the very beginning (from the field equations). The resultant cosmological equations are 
different from the ones obtained including $G(t)$ and $c(t)$ later, in the otherwise obtained ($G$ and $c$ constant) standard cosmological equations.}

In reference \cite{BelAl} time varying $G(t)$ and $c(t)$ have been introduced from the very beginning. Expanding relation (\ref{e2}) one obtains
\begin{equation}
\frac{\rho '}{\rho} + 3 (\omega +1) H + \frac{\Lambda ' c^4}{8 \pi G \rho} + \frac{G '}{G} - \frac{4 c'}{c} = 0
\label{e4}
\end{equation}

\noindent
where  $\rho$ is energy density, $\omega$ comes from the equation of state $p = \omega \rho$, $H$ is the Hubble parameter 
$H = a'(t)/a(t)$,  and $\Lambda$ is the cosmological constant. The primes in equation (\ref{e4}) refer to time derivatives. 
Taking $\omega$ and $\Lambda$ as constants, and using the condition (\ref{e1}), one immediately integrates equation (\ref{e4}) to obtain

\begin{equation}
\rho \ a ^{3 (\omega + 1)}  = \hbox{constant}
\label{e5}
\end{equation}

Using $E$ as the energy within a proper volume $V \propto a^3$ we transform equation (\ref{e5}) to
                          
\begin{equation}
E a ^{3 \omega }  = \hbox{constant}
\label{e6}
\end{equation}

Considering an initial radiation dominated universe $\omega = 1/3$, from equation (\ref{e6}) the energy is inversely proportional to $a(t)$. 
Then we get from equation (\ref{e6})

\begin{equation}
E =  \hbox{constant} / a(t) \approx N_{\gamma} \hbar c / \lambda
\label{e7}
\end{equation}

\noindent
where $N_{\gamma}$  is the number of photons (it turns out to be necessarily constant) with average wavelength $\lambda$. 

For a dust dominated universe, considering the galaxies as points of dust, we know that $\omega = 0$. 
This implies from equation (\ref{e6}) the well known conservation of energy
\begin{equation}
E = M c^2  = \hbox{constant}
\label{e8}
\end{equation}

The dark energy component (assumed to be given by the constant $\Lambda$), which may be considered as the {\it sea} where the galaxies are {\it floating}, 
corresponds to the measured value $\omega \approx -1$. 
Then we get from equation (\ref{e5}) that the dark energy density is constant. 
Hence the dark energy is $\propto a^3$. It is increasing with the proper volume. 
This necessarily implies that {\bf the universe must be filled with dark energy sources}, 
giving an expanding universe with constant dark energy density. This is a universe that reminds us of the well known 
(Steady State theory) Bondi-Gold-Hoyle (1948) model: 
here we have now the picture of an expanding universe with dark energy distributed sources 
(homogeneous and isotropic following the cosmological principle) resulting in a background of constant dark energy density. 
Instead of creation of mass our picture here corresponds to creation of dark energy, as given by a distribution of sources in space.

Finally we point out that the gravitational force $F$ between two {\it point} masses $M$ and $m$, given by Newton as

\begin{equation}
F = - \frac{GMm}{r^2}
\label{e9}
\end{equation}

\noindent
is the classical expression showing that $F$ is proportional to the product of the two masses involved, $M$ and $m$, where $r$ is the distance between them. 
With our condition (\ref{e1}) this can be replaced by the following proportionality

\begin{equation}
F \propto (M c^2) (m c^2) / r^2 = \hbox{constant} / r^2
\label{e10}
\end{equation}

Here we see the relevance of the change of physical properties in the Newtonian expression for the gravitational force: 
it is the relativistic energy of the masses the physical property responsible for gravity.

\section{Conclusions}

The presence of matter particles or photons does not change the gravitational concept: the gravitational force is always proportional to the product of the relativistic energies. Photons do gravitate. And the sources of gravity are relativistic energies, instead of the classical concept that gravity is due to the presence of mass. The general relativity statement that masses influence space-time, and that space time tells masses how to move, is here replaced by energies instead of masses.

The relation $G^{1/2} = c^2$ implies the possibility of a linkage between special relativity and gravitation.

Finally we arrive at the conclusion that the universe must be filled with dark energy sources, 
dark energy entering the universe and expanding it with a constant dark energy density.

\end{document}